\documentclass[11pt]{article}
\usepackage{moriond,epsfig}
\usepackage{vntex}
\usepackage[T5,T1]{fontenc}

\def\etal {{\it et al.}}
\def\lvpn{$8$}
\def\lvgap{$9$}

\def\al{\alpha}
\def\be{\beta}
\def\ga{\gamma}
\def\de{\delta}

\def\et{\eta}

\def\ka{\kappa}
\def\la{\lambda}

\def\si{\sigma}

\def\ps{\psi}

\def\Ga{\Gamma}

\def\Ps{\Psi}

\def\mn{{\mu\nu}}

\def\fr#1#2{{{#1} \over {#2}}}
\def\half{{\textstyle{1\over 2}}}
\def\quar{{\textstyle{1\over 4}}}

\def\frac#1#2{{\textstyle{{#1}\over {#2}}}}

\def\abs#1{\left|{#1}\right|}

\def\lsim{\mathrel{\rlap{\lower4pt\hbox{\hskip1pt$\sim$}}
    \raise1pt\hbox{$<$}}}
\def\gsim{\mathrel{\rlap{\lower4pt\hbox{\hskip1pt$\sim$}}
    \raise1pt\hbox{$>$}}}
\def\sqr#1#2{{\vcenter{\vbox{\hrule height.#2pt
         \hbox{\vrule width.#2pt height#1pt \kern#1pt
         \vrule width.#2pt}
         \hrule height.#2pt}}}}

\def\prt{\partial}

\def\etal{{\it et al.}}
\def\thpr{{these proceedings$^\dagger$}}

\def\pt#1{\phantom{#1}}

\def\ol#1{\overline{#1}}

\def\ivb#1#2{e^{#1}_{{\pt{#1}}#2}}
\def\uvb#1#2{e^{#1#2}}

\def\hul#1#2{h^{#1}_{{\pt{#1}}#2}}

\def\ab{\overline{a}{}}

\def\cb{\overline{c}{}}

\def\sb{\overline{s}{}}

\def\twiddle{\lower4pt\hbox{\hskip-0pt{$\widetilde{}$}}}
\def\m@th{\mathsurround=0pt}
\def\cmapstochar{\mathrel{\rlap{
  \lower0.1pt\hbox{\hskip-1.75pt{$\mapstochar$}}}
  \raise0pt\hbox{\hskip2.5pt{$\twiddle$}}}}
\def\notsimfill{$\m@th\cmapstochar$}
\def\scroodle#1{\vbox{\ialign{##\crcr\notsimfill\crcr
  \noalign{\kern-4pt\nointerlineskip}
   $\hfil\displaystyle{#1}\hfil$\crcr}}}

\def\cmapstocharbig{\mathrel{\rlap{
  \lower0.1pt\hbox{\hskip0.25pt{$\mapstochar$}}}
  \raise0pt\hbox{\hskip4.5pt{$\twiddle$}}}}
\def\notsimfillbig{$\m@th\cmapstocharbig$}
\def\scroodlebig#1{\vbox{\ialign{##\crcr\notsimfillbig\crcr
  \noalign{\kern-4pt\nointerlineskip}
   $\hfil\displaystyle{#1}\hfil$\crcr}}}

\def\mt{m^{\rm T}}

\def\af{(a_{\rm{eff}})}

\def\afb{(\ab_{\rm{eff}})}

\def\abt{(\ab^{\rm T}_{\rm{eff}})}
\def\abs{(\ab^{\rm S}_{\rm{eff}})}

\def\afbx#1{(\ab^{#1}_{\rm{eff}})}
\def\cbx#1{(\cb^{#1})}

\def\cbw{\cbx{w}}
\def\afbw{\afbx{w}}

\def\cs{(c^{\rm S})}

\def\cbt{(\cb^{\rm T})}

\def\atwt{(\scroodle{a}{}^{\rm T}_{\rm{eff}})}

\def\lrpartial{\raise 1pt\hbox{$\stackrel\leftrightarrow\partial$}}

\def\lrDmu{\stackrel{\leftrightarrow}{D_\mu}}

\def\a{$a_\mu$}
\def\bmu{$b_\mu$}
\def\cmn{$c_{\mu\nu}$}

\def\e{$e_\mu$}
\def\f{$f_\mu$}
\def\g{$g_{\la\mu\nu}$}
\def\H{$H_{\mu\nu}$}

\def\summarya{XIV}
\def\summaryb{XV}

\newcommand{\beq}{\begin{equation}}
\newcommand{\eeq}{\end{equation}}
\newcommand{\bea}{\begin{eqnarray}}
\newcommand{\eea}{\end{eqnarray}}
\newcommand{\bit}{\begin{itemize}}
\newcommand{\eit}{\end{itemize}}

\def\pno#1{PNO(#1)}

\begin{document}
\vspace*{4cm}
\title{LORENTZ SYMMETRY, THE SME, AND GRAVITATIONAL EXPERIMENTS}

\author{JAY D.\ TASSON}

\address{Department of Physics, Whitman College\\
Walla Walla, WA 99362, USA\\
E-mail: tassonjd@whitman.edu}

\maketitle\abstracts{
This proceedings contribution
summarizes the implications of
recent SME-based investigations of Lorentz violation
for gravitational experiments.}

\section{Introduction}

General Relativity along with the Standard Model of particle physics
provide a remarkable description of known physics.
Lorentz symmetry is a foundational principle of each,
and thus should be well tested experimentally.
It is also likely that General Relativity and the Standard Model
are limits of a more fundamental theory
that provides consistent predictions at the Planck scale.
Tests of  Lorentz symmetry provide a technically feasible means
of searching for potential suppressed signals from the Planck scale
in existing experiments and observations.\cite{ksp}
The gravitational Standard-Model Extension (SME)
provides a comprehensive test framework 
for searching for such potential signals
across all areas of known physics.\cite{akgrav,ck}

In spite of both the many high-sensitivity investigations 
of Lorentz symmetry \cite{tables}
performed in the context of 
the SME
in Minkowski spacetime,\cite{ck}
the scope of which continues to deepen and broaden,\cite{tables,nonmin,flat}
and the many investigations of metric theories of gravity
performed in the context of the
Parametrized Post-Newtonian (PPN) formalism,\cite{will}
there remain numerous potential Lorentz-violating deviations
from General Relativity
that have not yet been sought
observationally and experimentally.
Some of these violations
would lead to qualitatively new types of signals.

Lorentz-violating effects
in gravitational experiments
can originate from two basic places:
the pure-gravity action
and gravitational couplings in the other sectors
of the theory.
While some distinct theoretical issues are associated with each origin,
and some of the associated experimental signatures 
are quite different,
the relevant effects can be observed 
in many of the same classes of experiments.
The pure-gravity sector was the subject of Ref.\ \lvpn,
and Sec.\ \ref{pureg} summarizes some of the key theoretical issues
associated with that work.
Section \ref{mattg} provides a similar summary of theoretical issues
associated with matter-gravity couplings,
which were the subject of Ref.\ \lvgap.
Experiments relevant for investigations of Lorentz violation
originating from both sectors
are then considered in Sec.\ \ref{expt}.

\section{Lorentz violation in pure gravity}
\label{pureg}

Investigations of Lorentz violation performed in the context
of the  minimal SME
in Minkowski spacetime
are extended to include the post-Newtonian implications
of Lorentz violation in the pure-gravity sector
in Ref.\ \lvpn,
and several associated theoretical and phenomenological investigations
have expanded aspects of that work.\cite{gravthry,qbdoppler}
The action for the minimal pure gravity sector takes the form \cite{akgrav}
\beq
S = \fr {1}{2\ka} \int d^4x e (R - u R 
+s^\mn R^T_\mn + t^{\ka\la\mu\nu} C_{\ka\la\mu\nu}),
\label{llv}
\eeq
where $C_{\ka\la\mu\nu}$ is the Weyl tensor,
$R^T_\mn$ is the traceless Ricci tensor,
and $e$ is the vierbein determinant.
The first term here
is the standard Einstein-Hilbert term.
The relevant Lorentz-violating signals in the post-Newtonian analysis to follow
stem from the third term
involving the coefficient field $s^\mn$.
The second term is not Lorentz violating,
and the fourth term provides no contributions
in the post-Newtonian analysis.

It has been shown
that in the present context of Riemannian spacetime,
consistent Lorentz symmetry breaking must be spontaneous,\cite{akgrav}
though use of more general geometries may admit explicit breaking.\cite{akfin}
A number of implications stemming directly from 
spontaneous symmetry breaking
can also be considered;\cite{rbak,rpak,yb}
however,
a detailed discussion of these issues is beyond the scope
of the present discussion.

As a result of the specialization to spontaneous symmetry breaking
in the present context,
a primary theoretical issue addressed in Ref.\ \lvpn\
is establishing a procedure for correctly accounting for the fluctuations
in the coefficient fields,
including the massless Nambu-Goldstone modes 
of Lorentz-symmetry breaking.\cite{rbak}
This challenge is met in a general way,
without specializing to a specific model of spontaneous symmetry breaking,
under a few mild assumptions.
Upon addressing this issue,
the leading-order Lorentz-violating contributions
to the linearized field equations are obtained
and can be expressed in terms of the metric fluctuation $h_\mn$
and the vacuum value $\sb^\mn$ associated with the coefficient field $s^\mn$.
The post-Newtonian metric
is obtained from these equations.
With a suitable gauge choice \cite{lvpn}
the metric takes the form
\bea
g_{00} &=& -1 + 2U + 3 \sb^{00} U 
+\sb^{jk} U^{jk} - 4 \sb^{0j} V^j + O(4), 
\label{g00}\\
g_{0j} &=& -\sb^{0j}U - \sb^{0k} U^{jk} 
- \frac 72 (1 + \frac {1}{28} \sb^{00})V^j 
+\frac 34 \sb^{jk} V^k 
- \frac 12 (1+\frac {15}{4} \sb^{00})W^j
\nonumber\\
&&
+\frac 54 \sb^{jk} W^k
+\frac 94 \sb^{kl} X^{klj}
-\frac {15}{8} \sb^{kl} X^{jkl}
-\frac 38 \sb^{kl} Y^{klj},
\label{g0j}
\\
g_{jk} &=& \de^{jk} 
+ (2 - \sb^{00})\de^{jk} U
+ ( \sb^{lm} \de^{jk} 
- \sb^{jl} \de^{mk}
-\sb^{kl} \de^{jm}
+ 2\sb^{00} \de^{jl} \de^{km} ) U^{lm},
\label{gjk}
\eea
where $U$, $U^{jk}$, $V^j$, $W^j$, $X^{klj}$, and $Y^{klj}$
are potentials formed from appropriate integrals
over the source body.
The explicit form of the potentials is provided
in Ref.\ \lvpn.

The above metric is then compared and contrasted with the PPN metric.
The basic idea is that the pure-gravity sector of the minimal SME
provides an expansion about the action of General Relativity,
while the PPN provides an expansion about the metric.
Perhaps surprisingly,
an overlap of only one parameter is found
between the 20 coefficients in the minimal pure-gravity sector of the SME
and the 10 parameters of the PPN formalism.
This implies that leading corrections to General Relativity
at the level of the action
do not match
those typically studied in an expansion about the metric.
Note also
that the focus of the SME is on Lorentz violation throughout physics,
while the focus of the PPN is on deviations from General Relativity,
which may or may not be Lorentz violating.
Thus the minimal pure-gravity sector of the SME and the PPN formalism
provide complementary approaches to studying deviations from General Relativity.

Finally,
Ref.\ \lvpn\ introduces bumblebee models\cite{ks}
that provide specific examples of complete theories
with spontaneous symmetry breaking
that fit into the post-Newtonian results established
in the general context of the SME.

\section{Lorentz violation in matter-gravity couplings}
\label{mattg}

Though many high-sensitivity investigations 
of Lorentz violation \cite{tables}
have been performed in the context of 
the fermion sector of the minimal SME
in Minkowski spacetime,\cite{ck}
there remains a number of coefficients for Lorentz violation
in that sector that have not been investigated experimentally.
A methodology
for obtaining sensitivities to some of these open parameters
by considering gravitational couplings
in the fermions sector of the SME
is provided by Ref.\ \lvgap.
The set of coefficients $\ab_\mu$
for baryons and charged leptons,
which are unobservable in principle
in Minkowski spacetime,
is of particular interest.
Due to gravitational countershading,\cite{akjt}
these coefficients could be large 
relative to existing matter-sector sensitivities.

Prior to developing the necessary results
for experimental analysis,
the theoretical portion of Ref.\ \lvgap\
addresses a number of useful conceptual points.
One such point is consideration
of the circumstances under which relevant types of Lorentz violation
are observable in principle.
Though the $\ab_\mu$ coefficient
can be removed from the single fermion theory
in Minkowski spacetime
via a spinor redefinition,
it is highlighted that
it cannot typically be removed in the presence of gravity.\cite{akgrav}
This results in the gravitational countershading
pointed out in Ref.\ 17.

A coordinate choice
that can be used to fix the sector of the theory
that defines isotropy is also discussed,
and the role of the gravitational sector
in this context is established.
Ultimately,
the photon sector is chosen
to have $\et_\mn$ as the background metric,
though no generality is lost,
and other choices can be recovered.

The treatment of the fluctuations in the coefficient fields
established for the gravitational sector
is adapted to the context of matter-gravity couplings.
Two notions of perturbative order are introduced
to treat the fluctuations perturbatively
under the assumptions that gravitational and Lorentz-violating corrections
are small.
One notion of perturbative order, 
denoted O($m,n$),
tracks the orders in Lorentz violation and in gravity.
Here the first entry represents the order 
in the coefficients for Lorentz violation,
and the second entry represents the order in the metric fluctuation $h_\mn$.
A secondary notion of perturbative order,
which
tracks the post-Newtonian order,
is denoted \pno{$p$}. 
The O(1,1) contributions
are of primary interest
in Ref.\ \lvgap,
since the goal of that work is to investigate
dominant Lorentz-violating implications
in matter-gravity couplings.

To proceed toward the analysis of relevant experiments,
the results necessary for working at a number of energy levels
are developed from the full field-theoretic action
of the gravitationally coupled fermion sector of the SME,
which takes the form
\beq
S_\ps = 
\int d^4 x (\half i e \ivb \mu a \ol \ps \Ga^a \lrDmu \ps 
- e \ol \ps M \ps).
\label{fermion}
\eeq
where
\bea
\Ga^a
&\equiv & 
\ga^a - c_{\mu\nu} \uvb \nu a \ivb \mu b \ga^b
- d_{\mu\nu} \uvb \nu a \ivb \mu b \ga_5 \ga^b
\nonumber\\
&&
- e_\mu \uvb \mu a 
- i f_\mu \uvb \mu a \ga_5 
- \half g_{\la\mu\nu} \uvb \nu a \ivb \la b \ivb \mu c \si^{bc} 
\label{gamdef}
\eea
and
\beq
M
\equiv 
m + a_\mu \ivb \mu a \ga^a 
+ b_\mu \ivb \mu a \ga_5 \ga^a 
+ \half H_{\mu\nu} \ivb \mu a \ivb \nu b \si^{ab},
\label{mdef}
\eeq
where \a, \bmu, \cmn, $d_\mn$, \e, \f, \g, \H\ 
are coefficient fields for Lorentz violation.

Starting from Eq.\ \ref{fermion},
the relativistic quantum mechanics 
in the presence of gravitational fluctuations
and Lorentz violation is established
after investigating two methods of identifying
an appropriate hamiltonian.
The explicit form of the relativistic hamiltonian
involving all coefficients for Lorentz violation
in the minimal fermion sector is provided.

The standard Foldy-Wouthuysen procedure
is then employed to obtain the nonrelativistic quantum Hamiltonian.
At this stage,
attention is specialized to the study
of spin-independent Lorentz-violating effects,
which are governed by the coefficient fields $\af_\mu$, $c_\mn$
and the metric fluctuation $h_\mn$.
Though interesting effects may exist
in couplings involving spin, gravity, and Lorentz violation,
the pursuit of spin-independent effects
maintains a reasonable scope
focused on the least well-constrained coefficients
including the countershaded $\ab_\mu$ coefficients.

For many relevant applications,
the classical theory \cite{nr} 
associated with the quantum-mechanical dynamics
is the most useful description.
Thus the classical theory
involving nonzero $\af_\mu$, $c_\mn$, and $h_\mn$
is established at leading order in Lorentz violation 
for the case of 
the fundamental particles appearing in QED
as well as for bodies involving many such particles.
The modified Einstein equation
and the equation
for the trajectory of a classical test particle
follow from the classical theory.
Obtaining explicit solutions for the trajectories
of particles requires knowledge of the coefficient and metric fluctuations.
A systematic procedure
for calculating this information
is established,
and general expressions 
for the coefficient and metric fluctuations
are obtained to O(1,1)
in terms of gravitational potentials
and the vacuum values $\afb_\mu$ and $\cb_\mn$
associated with the coefficient fields
$\af_\mu$ and \cmn.
With this,
we find that the equation of motion
for a test particle
can be written
\bea
\ddot{x}^\mu &=&
- \Ga_{(0,1) \pt{\mu} \al \be}^{\pt{(0,0)} \mu} u^\al u^\be 
- \Ga_{(1,1) \pt{\mu} \al \be}^{\pt{(1,1)} \mu} u^\al u^\be
+ 2 \et^{\mu\ga} \cbt_{(\ga \de)} 
\Ga_{(0,1) \pt{\de} \al \be}^{\pt{(0,1)} \de} u^\al u^\be
\nonumber
\\
&&
+ 2 \cbt_{(\al \be)} 
\Ga_{(0,1) \pt{\al} \ga \de}^{\pt{(0,1)} \al} 
u^\be u^\ga u^\de u^\mu
- \fr 1 \mt
[\prt^\mu \atwt_\al - \et^{\mu\be} \prt_\al \atwt_\be ] 
u^\al,
\label{oogeo}
\eea
where the metric to be inserted into the Christoffel symbols is
\bea
g_{00} &=& - 1 +  2 \left[1
+ 2 \fr{\al}{m} \abs_0 + \cs_{00}\right] U 
+ 2 \left[\fr{\al}{m} \abs_j + 2 \cs_{(j0)}\right]V^j
- 2 \fr{\al}{m} \abs_j W^j,
\\
g_{0j} &=& \fr{\al}{m} \abs_j U
+ \fr{\al}{m} \abs_k U^{jk}
- \left[4 + \fr{\al}{m} \abs_0 + 4 \cs_{00}\right] V^j - \al \abs_0 W^j,
\\
g_{jk} &=& \de^{jk}
+ 2 \left[1 - \fr{\al}{m} \abs_0 + \cs_{00}\right] U \de^{jk}
+ 2 \fr{\al}{m} \abs_0 U^{jk}.
\label{metrica}
\eea
and the fluctuations in the coefficient field $\af_\mu$ take the form
\beq
\atwt_\mu^{(1,1)} = 
\half \al h_\mn \abt^\nu 
- \quar \al \abt_\mu \hul{\nu}{\nu} + \prt_\mu \Ps.
\label{aftw}
\eeq
Here the superscripts S and T indicate coefficients associated with the source
and test bodies
respectively,
and a dot over a quantity indicates a derivative with respect
to the usual proper time.
The subscripts on the Christoffel symbols indicate
the order in the small quantities that should be included in the given
Christoffel symbol.
The vacuum values $\afb_\mu$ and $\cb_\mn$
can then be identified with the coefficients for Lorentz violation
investigated in the Minkowski spacetime SME.

As in the pure-gravity sector,
bumblebee models provide specific examples
of the general results.

\section{Experiments}
\label{expt}

\subsection{Laboratory Tests}

The effects of coefficients $\afb_\mu$, $\cb_\mn$, and $\sb_\mn$
can be measured in a wide variety of experiments
performed in Earth-based laboratories.
Tests of this type that have been proposed or performed
include gravimeter experiments,
tests of the universality of free fall,
and experiments with devices 
traditionally used as tests of gravity at short range.

Analysis performed in Ref.\ \lvpn\ 
for the case of $\sb_\mn$,
and in Ref.\ \lvgap\
for $\afb_\mu$ and $\cb_\mn$,
reveals that the gravitational force acquires tiny corrections
both along and perpendicular to the usual free-fall trajectory
near the surface of the Earth.
Coefficients $\afb_\mu$ and $\cb_\mn$
also lead to a modified effective inertial mass of a test body
that is direction dependent,
resulting in a nontrivial relation between force and acceleration.
Both the corrections to the gravitational force
and to the inertial mass
are time dependent with variations
at the annual and sidereal frequencies.
In addition,
corrections due to 
$\afb_\mu$ and $\cb_\mn$
are particle-species dependent.

Based on the above discussion,
laboratory tests using Earth as a source
fall into 4 classes.
Free-fall gravimeter tests
monitor the acceleration
of free particles over time,
while force-comparison gravimeter tests
monitor the gravitation force on a body over time.
Both types of gravimeter tests
are sensitive to $\afb_\mu$, $\cb_\mn$, and $\sb_\mn$ coefficients.
The relative acceleration of,
or relative force on a pair of test bodies
can also be monitored
constituting free-fall and force-comparison Weak Equivalence Principle 
(WEP) tests
respectively.
Sensitivities to $\afb_\mu$ and $\cb_\mn$
can be achieved in WEP tests.
Relevant devices presently used for the above types of tests
include experiments with
falling corner cubes,\cite{fc}
atom interferometers,\cite{ai,aigrav,mh}
superconducting levitation,\cite{fcgrav}
tossed masses,\cite{tossed}
balloon drops,\cite{balloon}
drop towers,\cite{bremen}
sounding rockets,\cite{srpoem}
and
torsion pendula.\cite{tpwep}
Refs.\ \lvpn\ and \lvgap\
provide specific predictions
and estimated sensitivities for the above tests
including a frequency decomposition of the relevant signal
to which experimental data could be fit.
Note that the effective WEP violation
with periodic time dependence
considered here
is a qualitatively different signal that would likely have been missed
in past WEP tests.
One experiment of this type has already been performed
using an atom-interferometer
as a free-fall gravimeter.\cite{aigrav}

Variations of the above laboratory tests
involving the gravitational couplings 
of charged particles, antimatter,
and second- and third-generation particles
are also studied
in Ref.\ \lvgap.
Though they are very challenging experimentally,
these tests can yield sensitivities to Lorentz and CPT violation 
that are otherwise difficult or impossible to achieve.
Charged-particle interferometry,\cite{chargeai}
ballistic tests with charged particles,\cite{charge}
gravitational experiments with antihydrogen,\cite{anti}
and signals in muonium free fall \cite{muon}
are considered.
Some features of antihydrogen tests
are illustrated with simple toy-models limits
of the SME.

Though less sensitive at present
to the range-independent SME effects presently under discussion,
systems in which both the source mass 
and the test mass are contained within the lab,
such as those
devices 
traditionally used as tests of gravity at short range,
can also be considered.
A search for $\sb_\mn$ has been performed
using a cantilever system \cite{jl}
and a search for $\afb_\mu$
using a torsion-strip balance \cite{cspeake}
have been performed
using this approach.
A proposal to measure $\sb_\mn$
using a torsion pendulum with an asymmetric mass distribution
also exists.\cite{lvpn}

\subsection{Satellite-Based Tests}

Space-based experiments can offer unique advantages in testing
gravitational physics
\cite{spacegrav} 
and in searching for Lorentz violation.\cite{spaceexpt}
The WEP tests considered above are an example
of a class of tests
for which significant sensitivity improvements might be possible
in space,
due to the long free-fall times that may be attainable
on a drag-free spacecraft. 
There are several proposals for such missions
in the advanced stages of development, 
including
the Micro-Satellite \`a tra\^in\'ee Compens\'ee
pour l'Observation du Principe d'Equivalence
(MicroSCOPE),\cite{muscope}
the Satellite Test of the Equivalence Principle (STEP),\cite{step}
and the Galileo Galilei (GG) mission.\cite{gg}
A WEP experiment with reach similar to that of STEP
has also been suggested for the 
Grand Unification and Gravity Explorer
(GaUGE) mission.\cite{gauge}

Monitoring the relative motion
of test bodies of different composition
as they obit the Earth inside of the spacecraft
is the basic idea underlying these missions.
Nonzero coefficients for Lorentz violation
$\afbw_\mu$ and $\cbw_\mn$,
would result in material dependent orbits.
Ref.\ \lvgap\
provides the differential acceleration
of the test masses,
decomposed by frequency,
that are relevant for fitting data for each of the above proposed tests,
and achievable sensitivities are estimated.
As in the lab-based tests,
the SME signals would be distinguished from other sources of WEP violation
by the characteristic time dependences of the signals.
A ground-based version of the GG experiment,
Galileo Galilei on the Ground (GGG),\cite{gg}
which is presently taking data,
could also obtain sensitivities to Lorentz violation.

Another test with sensitivity to Lorentz violation
that was made possible using a space-based platform
is the gyroscope experiment, 
Gravity Probe B (GPB).\cite{gpb}
The geodetic or de
Sitter precession about an axis perpendicular to the
orbit and the gravitomagnetic frame-dragging or Lens-Thirring precession 
about the spin axis of the Earth
are the primary conventional relativistic effects for
a gyroscope in orbit around the Earth. 
An analysis of such a system in the presence of $\sb_\mn$
was performed in Ref.\ \lvpn.
It was found that additional Lorentz-violating precessions result,
including a precession about an axis perpendicular to both 
the angular-momentum axis of the orbit
and Earth's spin axis.
A similar investigation
considering the effects of $\afb_\mu$ and $\cb_\mn$
is possible based on the theoretical work
in Ref.\ \lvgap,
but it remains an open problem at present.

\subsection{Orbital Tests}

The search for anomalous effects
on orbits provides a natural way of testing gravitational physics.
References \lvpn\ and \lvgap\
consider tests that search for such effects
via laser ranging to the Moon and other bodies,
perihelion precession measurements,
and binary-pulsar observations.

Lunar laser ranging provides extraordinarily sensitive
orbital measurements.\cite{llr}
Based on the detailed proposal
to search for the effects of pure-gravity sector coefficient $\sb_\mn$
provided by Ref.\ \lvpn,
some of the best existing constrains on several components of that coefficient
have been placed using lunar laser ranging data.\cite{llrsme}
A similar proposal to search of $\afb_\mu$ and $\cb_\mn$ effects
on the lunar orbit is made in Ref.\ \lvgap.
Ranging to other satellites in different orientations
or of different composition
could yield additional independent sensitivities.

Measurements of the precession of the perihelion
of orbiting bodies \cite{peri} are also considered
for the case of $\afb_\mu$ and $\cb_\mn$ coefficients \cite{lvgap}
as well as $\sb_\mn$ coefficients.\cite{lvpn}
Based on the established advance of the perihelion
for Mercury and for the Earth, 
constraints on combinations of $\afb_\mu$, $\cb_\mn$, and $\sb_\mn$
are placed.
These constrains provide the best current sensitivity to $\afb_J$,
though it comes as a part of a complicated combination of coefficients.

Binary-pulsar observations complement the above solar-system tests
by providing orbits of significantly different orientations.\cite{pulsar}
Reference \lvpn\ contains detailed predictions
for the effects of $\sb_\mn$ on binary-pulsar systems.
The effects of $\afb_\mu$ and $\cb_\mn$
on such systems could also be investigated,
but detailed observational predictions remain
to be made.

\subsection{Photon and Clock Tests}

A final class of tests 
involves the interaction of photons with gravity
as well as effects on the clocks typically associated
with such tests.
References 11
and \lvgap\
consider signals arising 
in measurements of the time delay,
gravitational Doppler shift,
and gravitational redshift,
along with
comparisons of the behaviors of photons and massive bodies
for Lorentz violation in the pure gravity sector
and matter sector respectively.
Null redshift tests are also considered
in Ref.\ \lvgap\
resulting in expected sensitivity
to $\afb_\mu$ and $\cb_\mn$ coefficients.
Implications for a variety of existing and proposed experiments
and space missions are considered.\cite{photon}
An analysis of a variety of clocks
has been performed and sensitivities
to $\afb_\mu$ and $\cb_\mn$ coefficients have been achieved.\cite{mh}
Note that these results
and proposals are in addition to the Minkowski spacetime clock experiments
which have been performed on the ground \cite{tables}
and could be improved in space.\cite{spaceexpt}

\section{Summary}
 
Existing sensitivities from the experiments and observations
summarized above can be found in
{\it Data Tables for Lorentz and CPT Violation}.\cite{tables}
Expected sensitivities based on the proposals summarized above
are collected in
Table 6 of Ref.\ \lvpn\ and
Tables \summarya\ and \summaryb\
of Ref.\ \lvgap.
These sensitivities
reveal excellent prospects for using gravitational experiments
to seek Lorentz violation.
Of particular interest
are the opportunities to measure
the countershaded coefficients $\afb_\mu$
since
these coefficients typically cannot be detected 
in nongravitational searches.\cite{akjt}
Thus the tests of Lorentz symmetry proposed 
in Refs.\ \lvpn\ and \lvgap\
offer promising opportunities
to search for signals of new physics,
potentially originating at the Planck scale.
The effects can be sought in existing, planned, or feasible experiments
and in some cases provide experimental signatures that are
qualitatively different from those sought to date.

\noindent
\dag E.\ Aug\'e, J.\ Dumarchez, and J.\ Tr\^an Thanh V\^an, ed.,
{\it Proceedings of the XLVIth Rencontres
de Moriond and GPhyS Colloquium}, 
{\fontencoding{T5}\selectfont
Th\'\ecircumflex{} Gi\'\ohorn i, Vietnam, 2011}.


\section*{References}
\begin{thebibliography}{xx}

\bibitem{ksp}
V.A.\ Kosteleck\'y and S.\ Samuel,
Phys.\ Rev.\ D {\bf 39}, 683 (1989);
V.A.\ Kosteleck\'y and R.\ Potting,
Nucl.\ Phys.\ B {\bf 359}, 545 (1991).

\bibitem{akgrav}
V.A.\ Kosteleck\'y,
Phys.\ Rev.\ D {\bf 69}, 105009 (2004).

\bibitem{ck}
D.\ Colladay and V.A.\ Kosteleck\'y,
Phys.\ Rev.\ D {\bf 55}, 6760 (1997);
Phys.\ Rev.\ D {\bf 58}, 116002 (1998).

\bibitem{tables}
{\it Data Tables for Lorentz and CPT Violation,}
2010 edition,
V.A.\ Kosteleck\'y and N.\ Russell,
Rev.\ Mod.\ Phys.\ {\bf 83}, 11 (2011),
arXiv:0801.0287v4.

\bibitem{nonmin}
V.A.\ Kosteleck\'y and M.\ Mewes,
Phys.\ Rev.\ D {\bf 80}, 015020 (2009).

\bibitem{flat}
M.\ Nagel, K.\ M\"ohle, K. D\"oringshoff,
E.V.\ Kovalchunck, and A.\ Peters, \thpr;
M.\ Nagel and A.\ Peters, \thpr.

\bibitem{will}
C.M.\ Will,
{\it Theory and Experiment in Gravitational Physics},
Cambridge University Press, Cambridge, 1993.

\bibitem{lvpn}
Q.G.\ Bailey and V.A.\ Kosteleck\'y,
Phys.\ Rev.\ D {\bf 74}, 045001 (2006).

\bibitem{lvgap}
V.A.\ Kosteleck\'y and J.D.\ Tasson,
Phys.\ Rev.\ D {\bf 83}, 016013 (2011).

\bibitem{gravthry}
Q.G.\ Bailey,
Phys.\ Rev.\ D {\bf 82}, 065012 (2010);
B.\ Altschul, Q.G.\ Bailey, and V.A.\ Kosteleck\'y,
Phys.\ Rev.\ D {\bf 81}, 065028 (2010);
M.D.\ Seifert,
Phys.\ Rev.\ D {\bf 79}, 124012 (2009).

\bibitem{qbdoppler}
Q.G.\ Bailey,
Phys.\ Rev.\ D {\bf 80}, 044004 (2009).

\bibitem{akfin}
V.A.\ Kosteleck\'y,
Phys.\ Lett.\ B {\bf 701}, 137 (2011).

\bibitem{rbak}
R.\ Bluhm and V.A.\ Kosteleck\'y,
Phys.\ Rev.\ D {\bf 71} 065008 (2005);
R.\ Bluhm \etal, 
Phys.\ Rev.\ D {\bf 77}, 065020 (2008).

\bibitem{rpak}
V.A.\ Kosteleck\'y and R.\ Potting,
Gen.\ Rel.\ Grav.\ {\bf 37}, 1675 (2005);
Phys.\ Rev.\ D {\bf 79}, 065018 (2009). 

\bibitem{yb}
Y.\ Bonder, \thpr.

\bibitem{ks}
V.A.\ Kosteleck\'y and S.\ Samuel,
Phys.\ Rev.\ Lett.\ {\bf 63}, 224 (1989);
Phys.\ Rev.\ D {\bf 40}, 1886 (1989). 

\bibitem{akjt}
V.A.\ Kosteleck\'y and J.D.\ Tasson,
Phys.\ Rev.\ Lett.\ {\bf 102}, 010402 (2009).

\bibitem{nr}
V.A.\ Kosteleck\'y and N.\ Russell,
Phys.\ Lett.\ B {\bf 693}, 443 (2010).

\bibitem{fc}
I.\ Marson and J.E.\ Faller,
J.\ Phys.\ E {\bf 19}, 22 (1986);
K.\ Kuroda and N.\ Mio,
Phys.\ Rev.\ D {\bf 42}, 3903 (1990);
T.M.\ Niebauer, M.P.\ McHugh, and J.E.\ Faller,
Phys.\ Rev.\ Lett.\ {\bf 59}, 609 (1987).

\bibitem{ai}
A.\ Peters, K.Y.\ Chung, and S.\ Chu,
Nature {\bf 400}, 849 (1999);
J.M.\ McGuirk \etal, 
Phys.\ Rev.\ A {\bf 65}, 033608 (2002);
N.\ Yu \etal, 
Appl.\ Phys.\ B {\bf 84}, 647 (2006);
B.\ Canuel \etal,
Phys.\ Rev.\ Lett.\ {\bf 97}, 010402 (2006);
H.\ Kaiser \etal,
Physica B {\bf 385-386}, 1384 (2006);
S.\ Fray \etal, 
Phys.\ Rev.\ Lett.\ {\bf 93}, 240404 (2004).
S.\ Dimopoulos \etal, 
Phys.\ Rev.\ D {\bf 78}, 042003 (2008);
Y.-H.\ Lien \etal, \thpr.

\bibitem{aigrav}
K.-Y.\ Chung \etal, 
Phys.\ Rev.\ D {\bf 80}, 016002 (2009);
H.\ M\"uller \etal, 
Phys.\ Rev.\ Lett.\ {\bf 100}, 031101 (2008).

\bibitem{mh}
M.\ Hohensee \etal,
Phys.\ Rev.\ Lett.\ {\bf 106} 151102 (2011);
M.\ Hohensee and H.\ M\"uller, \thpr.

\bibitem{fcgrav}
R.J.\ Warburton and J.M.\ Goodkind,
Astrophys.\ J.\ {\bf 208}, 881 (1976);
S.\ Shiomi,
arXiv:0902.4081;
L.\ Carbone \etal,
in T.\ Damour, R.T.\ Jantzen, and R.\ Ruffini, eds.,
{\it Proceedings of the Twelfth Marcel Grossmann
Meeting on General Relativity},
World Scientific, Singapore, 2010.

\bibitem{tossed}
R.D.\ Reasenberg, 
in V.A.\ Kosteleck\'y, ed.,
{\it CPT and Lorentz Symmetry II}, 
World Scientific, Singapore, 2005.

\bibitem{balloon}
V.\ Iafolla \etal, 
Class.\ Quantum Grav.\ {\bf 17}, 2327 (2000).

\bibitem{bremen}
H.\ Dittus and C.\ Mehls,
Class.\ Quantum Grav.\ {\bf 18}, 2417 (2001).

\bibitem{srpoem}
R.D.\ Reasenberg and J.D.\ Phillips,
Class.\ Q.\ Grav.\ {\bf 27}, 095005 (2010).

\bibitem{tpwep}
Y.\ Su \etal, 
Phys.\ Rev.\ D {\bf 50}, 3614 (1994);
S.\ Schlamminger \etal, 
Phys.\ Rev.\ Lett.\ {\bf 100}, 041101 (2008);
T.\ Wagner, \thpr.

\bibitem{chargeai}
F.\ Hasselbach and M.\ Nicklaus,
Phys.\ Rev.\ A {\bf 48}, 143 (1993);
B.\ Neyenhuis, D.\ Christensen, and D.S.\ Durfee,
Phys.\ Rev.\ Lett.\ {\bf 99}, 200401 (2007).

\bibitem{charge}
F.S.\ Witteborn and W.M.\ Fairbank,
Phys.\ Rev.\ Lett.\ {\bf 19}, 1049 (1967).

\bibitem{anti}
G.\ Gabrielse,
Hyperfine Int.\ {\bf 44}, 349 (1988);
N.\ Beverini \etal, 
Hyperfine Int.\ {\bf 44}, 357 (1988);
R.\ Poggiani,
Hyperfine Int.\ {\bf 76}, 371 (1993);
T.J.\ Phillips,
Hyperfine Int.\ {\bf 109}, 357 (1997);
AGE Collaboration,
A.D.\ Cronin \etal,
{\it Letter of Intent: 
Antimatter Gravity Experiment (AGE) at Fermilab,}
February 2009;
D.\ Kaplan,
arXiv:1007.4956;
J.\ Walz and T.W.\ H\"ansch,
Gen.\ Rel.\ Grav.\ {\bf 36}, 561 (2004);
P.\ P\'erez \etal, 
{\it Letter of Intent to the CERN-SPSC,}
November 2007;
F.M.\ Huber, E.W.\ Messerschmid, and G.A.\ Smith,
Class.\ Quantum Grav.\ {\bf 18}, 2457 (2001);
AEGIS Collaboration,
A.\ Kellerbauer \etal,
Nucl.\ Instr.\ Meth.\ B {\bf 266}, 351 (2008);
C.\ Canali, \thpr;
P.\ Dupr\'e, \thpr.

\bibitem{muon}
K.\ Kirch,
arXiv:physics/0702143;
B.\ Lesche,
Gen.\ Rel.\ Grav.\ {\bf 21}, 623 (1989).

\bibitem{jl}
D.\ Bennett, V.\ Skavysh, and J.\ Long, in V.A.\ Kosteleck\'y, ed.,
{\it CPT and Lorentz Symmetry V}, 
World Scientific, Singapore, 2010.

\bibitem{cspeake}
H.\ Panjwani, L.\ Carbone, and C.C.\ Speake,
in V.A.\ Kosteleck\'y, ed.,
{\it CPT and Lorentz Symmetry V}, 
World Scientific, Singapore, 2010.

\bibitem{spacegrav}
For reviews of space-based tests of relativity see,
for example
C.\ L\"ammerzahl, 
C.W.F.\ Everitt, 
and F.W.\ Hehl, eds.,
{\it Gyros, Clocks, Interferometers \ldots :
Testing Relativistic Gravity in Space},
Springer, Berlin, 2001.

\bibitem{spaceexpt}
R.\ Bluhm \etal,
Phys.\ Rev.\ Lett.\ {\bf 88}, 090801 (2002);
Phys.\ Rev.\ D {\bf 68}, 125008 (2003).

\bibitem{muscope}
P.\ Touboul, M.\ Rodrigues, G.\ M\'etris, and B.\ Tatry,
Comptes Rendus de l'Acad\'emie des Sciences, Series IV,
{\bf 2}, 1271 (2001);
P.\ Touboul, R.\ Chhun, D.\ Boulanger,
M.\ Rodrigues, and G.\ Metris, \thpr;
A.\ Levy, P.\ Touboul, M.\ Rodrigues,
A.\ Robert, and G.\ Metris, \thpr.

\bibitem{step}
T.J.\ Sumner \etal,
Adv.\ Space Res.\ {\bf 39}, 254 (2007).

\bibitem{gg}
A.M.\ Nobili \etal,
Exp.\ Astron.\ {\bf 23}, 689 (2009);
A.M.\ Nobili \etal, \thpr.

\bibitem{gauge}
G.\ Amelino-Camelia \etal,
Exp.\ Astron.\ {\bf 23}, 549 (2009).

\bibitem{gpb}
C.W.F.\ Everitt \etal,
Phys.\ Rev.\ Lett.\ {\bf 106}, 221101 (2011).

\bibitem{llr}
J.G.\ Williams, S.G.\ Turyshev, and H.D. Boggs, 
Phys.\ Rev.\ Lett.\ {\bf 93}, 261101 (2004);
T.W.\ Murphy \etal, 
Pub.\ Astron.\ Soc.\ Pac.\ {\bf 120}, 20 (2008).

\bibitem{llrsme}
J.B.R.\ Battat, J.F.\ Chandler, and C.W.\ Stubbs, 
Phys.\ Rev.\ Lett.\ {\bf 99}, 241103 (2007).

\bibitem{peri}
C.M.\ Will,
Living Rev.\ Rel.\ {\bf 4}, 4 (2001).

\bibitem{pulsar}
I.H.\ Stairs, Living Rev.\ Rel.\ {\bf 6}, 5
(2003).

\bibitem{photon}
B.\ Bertotti, L.\ Iess, and P.\ Tortora,
Nature {\bf 425}, 374 (2003);
T.\ Appourchaux \etal, 
Exp.\ Astron.\ {\bf 23}, 491 (2009);
L.\ Iess and S.\ Asmar,
Int.\ J.\ Mod.\ Phys.\ D {\bf 16}, 2117 (2007);
P.\ Wolf \etal,
Exp.\ Astron.\ {\bf 23}, 651 (2009);
B.\ Christophe \etal, 
Exper.\ Astron.\ {\bf 23}, 529 (2009);
S.G.\ Turyshev \etal, 
Int.\ J.\ Mod.\ Phys.\ D {\bf 18}, 1025 (2009);
S.B.\ Lambert and C.\ Le Poncin-Lafitte,
Astron.\ Astrophys.\ {\bf 499}, 331 (2009);
R.\ Byer,
{\it Space-Time Asymmetry Research},
Stanford University proposal,
January 2008;
L.\ Cacciapuoti and C.\ Salomon,
Eur.\ Phys.\ J.\ Spec.\ Top.\ {\bf 172}, 57 (2009);
S.G. Turyshev and M.\ Shao,
Int.\ J.\ Mod.\ Phys.\ D {\bf 16}, 2191 (2007);
S.C.\ Unwin \etal,
Pub.\ Astron.\ Soc.\ Pacific {\bf 120}, 38 (2008);
M.\ Gai, \thpr;
N.\ Ashby and P.\ Bender, \thpr;
L.\ Cacciapuoti and C.\ Salomon, \thpr;
T.\ Schuldt \etal, \thpr.

\end{thebibliography}
\end{document}